\documentclass[11pt,a4paper]{article}
\usepackage{amsmath}
\usepackage{amsfonts}
\usepackage{amssymb}
\usepackage{inputenc}
\usepackage{times}
\usepackage{graphicx}

\newtheorem{axiom}{Theorem}[section]
\newtheorem{proposition}{Proposition}[section]
\newtheorem{lemma}{Lemma}[section]

\newtheorem{Appendix}{APPENDIX}[section]
\newtheorem{definition}{Definition}[section]
\newtheorem{remarks}{Remarks}[section]
\newtheorem{remark}{Remark}[section]
\begin{document}

\bibliographystyle{unsrt}

\def\boxit#1#2{\setbox1=\hbox{\kern#1{#2}\kern#1}%
\dimen1=\ht1 \advance\dimen1 by #1 \dimen2=\dp1 \advance\dimen2 by
#1
\setbox1=\hbox{\vrule height\dimen1 depth\dimen2\box1\vrule}%
\setbox1=\vbox{\hrule\box1\hrule}%
\advance\dimen1 by .4pt \ht1=\dimen1 \advance\dimen2 by .4pt
\dp1=\dimen2 \box1\relax}

\def\build#1_#2^#3{\mathrel{\mathop{\kern 0pt#1}\limits_{#2}^{#3}}}
 \def\K{\Bbb K}
\def\C{\Bbb C}
\def\R{\Bbb R}
 \def\N{\Bbb N}
 \def\I{\Bbb I}
\def\ecart{\noalign{\medskip}}
\font\tenmsb=msbm10 \font\sevenmsb=msbm7 \font\fivemsb=msbm5
\newfam\msbfam
\textfont\msbfam=\tenmsb \scriptfont\msbfam=\sevenmsb
\scriptscriptfont\msbfam=\fivemsb
\def\Bbb#1{{\fam\msbfam\relax#1}}
\vfill\eject
{\title{  The ``non triviality''  of a  $\Phi_4^4$ model,\ \ \ III\\
the ``Osterwalder-Schrader Positivity'' }
\author{
  \ Marietta Manolessou\\
  \textbf{CYTech}\,-\, Department of Mathematics}
\maketitle

\begin{abstract} 
The present paper, III, is the third part of a series of papers, under the global title  ``the  non triviality  of a  $\Phi_4^4$ model''. Parts I and II have been previously completed. In them thanks to the properties we dubbed ``splitting -tree structure'', and ``alternating signs'', which characterize our connected Green's functions,
 we have constructed   a unique non trivial solution to a  $\Phi_4^4$  non linear renormalized system of equations of motion in Euclidean space.

 In the  present work, we show how, by application of these properties, the solution of our $\Phi^4_4$ model  verifies the Osterwalder-Schrader Positivity requirement. This result complements those obtained in I and II where, apart from the Positivity, the Axiomatic Quantum Field theory properties have been established.
 
The O.S. Positivity is verified under a condition on the physical coupling constant relatively weaker than the one imposed in order to obtain the convergence of the  $\Phi^4_4$ mapping to the unique non trivial solution. 
   \end{abstract} 
\vfill\eject

\tableofcontents

\vfill\eject

\section{Introduction}
\pagenumbering{arabic}
\subsection{The verification by the $\Phi^{4}_4$ solution of the  Relativistic and Euclidean Quantum Field Theory Axioms }  

The subject of verification of the axioms  of Q.F.T in Minkowski  \cite{Wightman}\cite{St.W.},\cite{(Q.F.T.)}  or Euclidean \cite{O.S.1} \cite{O.S.2}    space by various physical interaction models  between  elementary particles    
has a long history. 

The litterature concerning the construction of quantum field theory models   with Euclidean Green's functions characterized by analogous  features with those verified by the Wightman functions, began already in the 50's. 

In \cite{Schwi} J. Schwinger, presented
the four-dimensional Euclidean formulation of Quantum Field Theory during the Annual International Conference on High Energy Physics at CERN,
  and published in the Proceedings of the National Academy of Sciences in 1958.

Later K. Symanzik in
\cite{Sym 0} proposed models in the Euclidean Quantum field theory context.
 
 The equivalence between the axioms of Q.F.T. in Minkowksi and Euclidean space 
 appears in its most rigorous form during the years 1973-75  with the works of
K. Osterwalder - R. Schrader \cite{O.S.1} \cite{O.S.2} \cite{Osterwalder},
 V. Glaser 
\cite{Glaser}, E. Nelson  \cite{Nelson}. and J. Fr\"olich in
\cite {Frolich}.

By the end of the seventies and until recently many papers did appear  concerning the particular Axiomatic Q.F.T. property of Positivity  under the name of  ``O.S. Positivity'' and more frequently of ``Reflection Positivity'' 
(cf. J. Glimm and A. Jaffe, in\cite{GlimmJaffe} and  
J. Fr\"olich in \cite {Frolich}). 
In particular we would like to refer the reader to the opening talk of A. Jaffe  \cite{Jaffe0} 
  ``Reflection Positivity Then and Now'' at the conference   dedicated to the memory of R. Schrader on November 20, 2017  held at the Mathematical Research Institute, Oberwolfach, Germany. The author not only expresses his enthousiasm for the discovery of the ``O.S. recontruction theorem'' but he points out how the principle of ``reflection positivity'' plays a crucial role in many domains of mathematical physics (cf. \cite{Jaffe1}). 
 
In the present work, the verification of the Osterwalder-Schrader axioms by  our $\Phi^4_4$ solution completes  our program of I \cite{MM7} and II \cite{MM8} towards the construction of
 an Axiomatic  Q.F.T. \cite{(Q.F.T.)} model. 

Briefly, 
 in \cite{MM}  starting from the equation of motion 	
and inspired by Zimermann's work \cite{Zim}, we introduced the ``Renormalized Normal product''
and established an equivalent infinite dynamical system of equations of motion
in ``four dimensions'' for the Green's functions (the 
``vacuum expectation  values'' of the theory)
  which  has the form reminded in Appendix \ref{Ap.4.4}.
  
Now, we complete  the results that we  established   previously partially in ref.\cite{ MDu} for the renormalized equations of motion,
  recently in \cite{MM2} and more precisely in \cite{MM7}, \cite{MM8} for the solution of these $\Phi^4_4$ equations of motion. 
  
  As a matter of fact, the linear  
 Axiomatic Q.F.T. properties together with the distribution property, Euclidean covariance and symmetry together with the linear Axiomatic Q.F.T. analyticity properties (in complex Minkowski space) related to the locality, spectrum and uniqueness of the vacuum (associated to the cluster property) have been established for our $\Phi^4_4$ model.   
 
 These results ensured in some sense the coherence of our scheme but not completely.  
 As far as the positivity property is concerned, the
 complete results in four
 (and automatically in all 
smaller) dimensions constitute the purpose of the present paper.
 
More precisely we prove that under  the \emph{``weak condition'' }($\Lambda<1/6$) imposed on the physical coupling constant the infinite
 sequence of Green's functions
whose connected part is the
 solution  of the
$\Phi^4_4$ equations of
 motion in Euclidean momentum space in \cite{MM7} and \cite{MM8},
 verifies the set
 of Osterwalder-Schrader Positivity
 Axioms (O.S.P) \cite{O.S.2}.

 In this way we ensure
 that the infinite sequence-solution is no longer
 formal but in view of the 
\emph{reconstruction theorem} (cf.\cite{St.W.}, and \cite{O.S.2}),
 it is a well defined infinite
 sequence of Green's functions
 equivalent to a nontrivial
 Wightman $Q.F.T.$
\begin{remark}\label{Rem1.1} We point out that by saying ``weak'' condition ($\Lambda<1/6$) imposed on the coupling constant, we  simply mean that it is relatively less restrictive than the  conditions we imposed in order to obtain the local contractivity ($\Lambda<0.04$) and the corresponding to the stability of $\Phi_4^4$-iteration $(\Lambda<0.05$) for the construction of the $\Phi_4^4$ non trivial solution obbtained in \cite{MM7} and \cite{MM8}.
\label{1.1}
\end{remark} 
  
\subsection{The equivalence between Q.F.T. Axioms  in Euclidean space  and the corresponding  in Relativistic  Minkowski Space }
\subsubsection{The Chart of Osterwalder-Schrader} 	 
Let us  remind that the main theorem proved in \cite{O.S.1} or \cite{O.S.2} is represented by the following chart of equivalences which connects the Euclidean Axioms of Osterwalder-Schrader and the Relativistic Wightman Axioms \cite{St.W.} \begin{equation}\begin{array}{l}  \ \ \ \    
\hspace{2.0cm}\mbox{\underline{ EUCLIDEAN}} \hspace{4.1cm} \mbox{\underline{ RELATIVISTIC}} \\
\ \ \ \qquad (Euc.1)\equiv\begin{pmatrix}\mbox{Temperedness}\cr
 \mbox{Covariance}\cr
\mbox{Positivity}\cr
\end{pmatrix} \hspace{0.5cm}\Longleftrightarrow\hspace{0.5cm}(Rel.1)\equiv\begin{pmatrix}\mbox{Temperedness}\cr
 \mbox{Covariance}\cr
\mbox{Positivity}\cr
\mbox{Spectrum}
\end{pmatrix} \\
(cf. \cite{O.S.1})\ \qquad \ \ \ \ \ \      \\
\ \ \ \qquad (Euc.1)\ \   +\ \quad \begin{pmatrix}\mbox{Symmetry}
\end{pmatrix}
 \hspace{0.8cm}\Longleftrightarrow\hspace{0.6cm}(Rel.1)\ \ +\ \ \begin{pmatrix}\mbox{Locality}\end{pmatrix} \\
 \ \ \ \\
 \ \ \ \qquad  (Euc.1)\ \ \  +  \quad \begin{pmatrix}\mbox{Cluster}
\end{pmatrix}
 \hspace{1.3cm}\Longleftrightarrow\hspace{0.6cm}(Rel.1)\ \ +\ \ \ \begin{pmatrix}\mbox{Cluster}\end{pmatrix} \\
 \ \ \  \\
 \end{array} 
 \label{1.1}
 \end{equation}

\subsubsection{Plan of the paper}	 
In the next section we recall the definition of
 the  O.S.P. conditions in x-Euclidean space,
 and present the analogous
 expression in terms of the non connected 
 Green's functions (time order product's expectation values in our formalism).
We then express them in terms of connected components.

By application of the Fourier
 transform together with the 
symmetry properties and Euclidean invariance, we
 reformulate the positivity in terms of the so called  $O.S.P.n$  conditions
 in the Euclidean four momentum space in terms of our Green's functions sequences, namely
 truncated (connected) completeley amputated wth respect to the free propagators Green's functions. 
 
  We complete  this section by   an auxillary lemma which represents the starting point  of the recursion used in the proof of the main theorem presented in section 3.
 
In the third section we establish the $O.S.P.n$ conditions in momentum space
for the non connected and connected part contributions. We first present two
auxiliary lemmas and then the theorem 3.1 which yields as corollary the main result theorem 3.2.

In the appendices   we give the detailed proofs
 of our statements together with some necessary reminders from \cite{MM7}\cite{MM8}.
 
 The basic tools of the proof are again the
 \emph{''alternating signs''} 
and the \emph{''splitting'' 
or factorization} properties of the Green's  
 functions in terms of ``tree type'' functions established
 previously 
in all dimensions $r$ with $0\leq r\leq4$ and at every value of
 the external momenta. 

As a matter of fact the signs and the
``tree type splitting'' (or factorization) properties 
of the connected Green's functions provide
the possibility to obtain another   decomposition 
 of the non connected (non truncated) Green's  function $\tau^{ n+1}$ in terms of its connected parts. This decomposition is different but equivalent to the ``classical'' one of definition \ref{Def.2.1} (cf. equation \ref{classicalDec} reminded later in section 3), and we  present it by Lemma \ref{3.2}.
 As a matter of fact it results from the successive application   
of the ``tree type'' decomposition $C^{n+1}/(-6\Lambda)$.

\vspace{8mm}

\section{ The  O.S.P.  conditions in momentum space}
\subsection{In $x$- space}
\vspace{5mm}

In \cite{O.S.1}\cite{O.S.2} the following conditions have been established 
by Osterwalder - Schrader in the
 Euclidean $x$ - space.
\begin{equation}
\sum_{\scriptscriptstyle M, N}{\cal G}_{\scriptscriptstyle (M+N)}
(\Theta \, g_{\scriptscriptstyle M}^*\,\times 
g_{\scriptscriptstyle N})\,  \geq\,  0 
\label{2.2}
\end{equation}

	
Where ${\cal G}$ means the Schwinger functions \cite{Sch} (distributions)
in Euclidean  $x$-space and it corresponds to the Wightman distributions in Minkowski space.

$g_{\scriptscriptstyle M}$
belongs to the space of test functions 
${\cal S}(\Bbb R^{4M}),\, 
(g_{\scriptscriptstyle M}^*$\, 
 means complex conjugate
 of $g_{\scriptscriptstyle M}$)
and 
$$(\Theta g)_{\scriptscriptstyle M}(x_1,\dots, x_M)=
g_{\scriptscriptstyle M}(\vartheta x_1\dots,\,
 \vartheta x_M).$$ 
 where for every vector $x=\{x^0,\, \vec x\}
\in \Bbb R^4\, :\, \vartheta x= \{-x^0,\, \vec x\} $.

\vspace{3mm}
In all that follows we denote by $\tau^{n+1}$ the Fourier transform (in the sense of distributions) in $q$-space  of the tempered distribution ${\cal G}$.      The connected (completely amputated with respect to the free propagators) parts of $\tau^{n+1}$, correspond (following our prescriptions) to the  $H^{n+1}$ Green's functions solutions of the equations that we introduced and studied  in \cite{MM1} \cite{MM2}, \cite{MM7} and\cite{MM8}. \ 

\subsection{The O.S.P.n  conditions in q-space  of non connected Green's functions $\tau^{ n+1}$}

By application of the isomorphisms of Fourier transform and its inverse on the product space of test functions ${\cal S}(\Bbb R^{4M})\times {\cal S}(\Bbb R^{4N}),$
\begin{equation}
{\cal F} : \ {\cal S}(\Bbb R^{4M})\times {\cal S}(\Bbb R^{4N}) \to {\cal S}(\Bbb R^{4M})\times {\cal S}(\Bbb R^{4N})
\label{Fourier}
\end{equation}
 we directly obtain the corresponding positivity
 conditions for the non truncated  Green's
 functions  (or time order product) in $q$-space 
 (momentum space).

$$\forall\  \ n=2k+1,\   \ k\in \Bbb N\ \ $$
\begin{equation}\begin{array}{l}
 \displaystyle{\sum_{1\leq M\leq n,
1\leq N\leq n\atop   M+N\leq n+1}} \int 
\tau^{ n+1}
(q_{{\scriptscriptstyle (n+1)}}) 
 \delta (Q_{{\scriptscriptstyle n+1 }})
\overline{\hat f_{\scriptscriptstyle {(M)}}
(q_{{\scriptscriptstyle (M)}})}\
\hat f_{\scriptscriptstyle {(N)}}(  q_{{\scriptscriptstyle (N)}}) dq_{{\scriptscriptstyle (M)}}dq_{{\scriptscriptstyle (N)}}   
  \geq 0\\
q_{{\scriptscriptstyle (M+N)}}=q_{{\scriptscriptstyle (n+1)}}=q_{{\scriptscriptstyle (M)}}
\cup  q_{{\scriptscriptstyle (N)}};\   q_{{\scriptscriptstyle (M)}}
\cap  q_{{\scriptscriptstyle (N)}}=\{q_{\scriptscriptstyle M}\}\subset q_{{\scriptscriptstyle (M)}},\     q_{\scriptscriptstyle M}=q_{\scriptscriptstyle 1} \in q_{{\scriptscriptstyle (N)}}\\
\mbox{and}\ \  Q_{{\scriptscriptstyle n+1 }}=\displaystyle\sum_{i=1}^{n+1} q_i\end{array}
\label{2.4}
\end{equation}

\begin{remarks}

\begin{enumerate}
 
\item{}
 Here $\hat f$ means the Fourier transform of an arbitrary test function  $f\in {\cal S}(\R^{4n})_x$. 
\item{} Following \cite{O.S.2} or equivalently \cite{St.W.}
when $n\to \infty$ the  conditions \ref{2.2} in x-space (or \ref{2.4} in q-space) ensure the positivity of the norm of every infinite dimensional vector of test fuctions $\{\hat f_n\}_{n \in \N}$, associated with the hermitian form (scalar product) given in terms of the tempered distribution $\tau$.
\item{}
Note that the Euclidean-translation invariance in $x-space$
 leads to the total
 energy momentum
conservation which is expressed by the $``\delta-function''$
 $\delta (Q_{{\scriptscriptstyle n+1}})$ appearing in the above formula.

 \item{}
 As we noticed before, the above form of the  O.S.P.   conditions are not
 suitable to be studied by our method
 because the characteristic bounds, signs, splitting, and  tree structure    properties of the
 $\Phi_4^4$ solution established in \cite{MM7}-\cite{MM8} and recalled in the Appendix \ref{Ap.4.4} are expressed
 in terms of the truncated or  connected and completely amputated with respect to the free propagators Green's functions $H^{n+1}$ . 

 Therefore, taking into account the
 decomposition formula in connected parts of every inverse
 Fourier transform of the
 $\tau^{ n+1}-function$   in   $x$-space  
and then by application of:
\begin{description}
\item[a.] the isomorphisms of Fourier transform and its inverse on the product space of test functions ${\cal S}(\Bbb R^{4M})\times {\cal S}(\Bbb R^{4N})$,
\item[b.] the \emph{symmetry} and \emph{Euclidean\  \ invariance}
 of every connected Green's
function in $x$-space, 
\end{description}
 we obtain in  a more
 appropriate expression of the  O.S.P.  conditions
 in Euclidean momentum $q$-space. We also notice that 
 we shall use the notation $(O.S.P.n)$ for reference either
 to the above set of inequalities \ref{2.4} (non connected expression) 
 or to the following (connected expressions) \ref{classicalDec} or \ref{2.9}.  
 \item{} Moreover, the fact that every connected part Green's function $H^{n+1}$ is a uniquely defined tempered distribution in the space ${\cal S'}(\Bbb R^{4n})$ as solution of the equations of motion
and continuous with repect to each one of its arguments, the other being constant, we are allowed to apply the Schwartz-Nuclear Theorem \cite{Schwartz} and target all the proofs which follow  to test functions which belong to the dense subset (of ${\cal S}(\Bbb R^{4M})\times {\cal S}(\Bbb R^{4N})$) of all linear combinations of the tensor product functions, namely:
 \begin{definition}   {(Factorization of the test functions)}  \label{def.factor} 
 \begin{equation}\begin{array}{l} 
  \tilde f_{\scriptscriptstyle (N)}\in {\cal S}(\R^4) \times {\cal S}(\R^4) \dots \dots {\cal S}(\R^4)  \\
 \tilde f_{\scriptscriptstyle (N)}(q_{{\scriptscriptstyle (N)}})=\displaystyle{\prod_{1\leq l\leq N} f_1^{(l)}(q_l)}
 \end{array}
 \label{Factor.}
 \end{equation}
 \end{definition}
\emph{ Notice that in the following for simplicity we often omit  the subscript $1$ from $ f_1^{(l)}$  and write $f_{\scriptscriptstyle (N)}$ instead of $  \tilde f_{\scriptscriptstyle (N)}$}.
\end{enumerate}    
\end{remarks}

\subsection{The $(O.S.P.n)$ conditions in q- space for the connected $H^{n+1}$ Green's functions)}
\begin{definition}
  $\forall \ n=2j+1, \ j \in\N$  we consider the set of odd positive  integers indices:
 \begin{equation} 
 (n)=\{1, 3, 5\dots \ , n\}
 \label{2.6}
 \end{equation}
 We introduce the set
   $\varpi_{n}$  of all partitions of  
 $(n)$  as follows: 
\begin{equation}\begin{array}{l}
\mbox{ A sequence $J$ of non empty disjoint subsets of $(n)$ belongs to $\varpi_{n}$,\ \  if:}\\
   J= (J_1,   J_2,\dots  J_k)  \ \  k\leq n  \ \ \ \mbox{and}\\
\forall\  i\in (1, 2,\dots k) ,\ \  \ j\in (1, 2,\dots k), \mbox{with} \  i\neq j \  J_i\cap J_j=\emptyset, \ \ \bigcup_{1\leq l\leq k} J_l=(n).\\ 
 \mbox{Moreover} \ \ {\rm Card}{J}_l =j_l, \mbox{where  $j_l,  \  l\in(1, 2,\dots k), $  are odd integers such that}, \\
 j_l\geq j_2\geq\dots j_{k-1}\geq j_k , \ \mbox{and}\ \displaystyle{\sum_{l=1}^{k}j_l=n}
  \end{array}
  \label{2.7}
\end{equation}
In the particular case $k=3$ we often use the notation 
 $\varpi_{n}(3)$ for the set of partitions $I= (I_1,   I_2,  I_3)$  with ${\rm Card}{I}_l =i_l $ where  $i_l, \  l \in (1, 2, 3)$ are odd integers such that:
$ i_1\geq i_2\geq i_3$ and $\displaystyle{\sum_{l=1}^{3}i_l=n}$
\end{definition}

\begin{definition} \textbf{(Connected parts' form of the $(O.S.P.n)$ conditions 
 and matrix representations)}  \label{Def.2.1}

We first  consider the standard decomposition of the non connected time order product in terms of the connected parts (and resp.connected completely amputated with respect to the free propagators): 
\begin{equation}\begin{array}{l}
\tau^{ n+1}(q_{{\scriptscriptstyle (n+1)}})=
\displaystyle{\sum_{J\in\varpi_{  n}} C_{ ( j_1, \dots, j_k)} 
 \prod_{1\leq l\leq k}\eta^{j_{l}+1}(q_{(j_{l}+1)}) \ 
\delta (Q_{j_{l}+1}) }\\
(\mbox{or respectively})\\
\tau^{ n+1}(q_{{\scriptscriptstyle (n+1)}})=\displaystyle{\sum_{J\in\varpi_{  n}} C_{ ( j_1, \dots, j_k)} 
 \prod_{1\leq l\leq k}H^{j_{l}+1}(q_{(j_l+1)})\prod_{1\leq r\leq j_{l}} \Delta_F(  q_r) \delta (Q_{j_{l}+1}) }\\
\mbox{here:} \   C_{ ( j_1, \dots, j_k)} 
= \displaystyle{\frac{n!}{j_1!\dots j_{k-1}! j_k! } }
 \end{array}
\label{classicalDec}
  \end{equation}

So equivalently with \ref{2.4} we have to ensure that, for every $n=2r+1, r \in \N$   and    $\forall\ (q,\Lambda) \in  
{\cal E}^{4n}\times \Bbb R^+$,
\begin{equation}\begin{array}{l}
\displaystyle{\sum_{1\leq N\leq {n},1\leq M\leq {n}\atop   N+M\leq n+1}}\displaystyle{\sum_{J\in\varpi_{n}}\int \overline{  {f}_{ (M)} }
\prod_{1\leq l\leq k}H^{j_{l}+1}}\displaystyle{\prod_{1\leq r\leq j_{l}}\Delta_F(q_r) {f}_{ (N)}dq_{(n)}}\geq 0 
\end{array}
\label{2.9}
\end{equation}
\end{definition}

 Here  $f_{(M)} $( (resp.$f_{(N)})$ are the factorized  test functions defined on the corresponding cartesian products of euclidean momentum spaces  as introduced before by  \ref{Factor.}.
 
 Moreover the $``\delta-function"$
 $\delta (Q_{j_{l}+1})$ (which appears  in \ref{classicalDec} for every connected part and expresses the total energy momentum conservation,  resulting from the Euclidean-translation invariance in $x$-space)  has disappeared in \ref{2.9} after the integration (Fubini) witn respect to every ``last'' momentum variable:
$$
 q_{j_{l}+1}=-\sum_{i=1}^{l}q_{j_{i}}
$$
and $dq(n)$ is an abbreviated notation for the Euclidean measure: 
$$dq(n)=\prod_{1\leq m\leq n}dq_{m}$$
in the space of $n$ independent momentum variables.
Finally we notice that
 often we simplify the notation of the arguments for the set of $n$ independent moments and right $(q)$ instead of $q_{\scriptscriptstyle (n)}$
 
 Notice that every term of the sum in \ref{2.4} (resp. of   double sum \ref{2.9}) is a hermitean form that can be represented as an element of a matrix representation as  in  the examples of \ref{matrices}  (given in Appendix \ref{App.4.1}).

 For  practical raisons we shall often  use a three parts  decomposition of  $ \tau^{ n+1}$:
\begin{equation}
 \tau^{ n+1}= T^{n}_1+ T^{n}_2 +T^{n}_3 \qquad
 \end{equation}
 where: 
 \begin{equation}\begin{array}{l}
  T^{n}_1= H^{n+1}\displaystyle{\prod_{l=1}^n\Delta_F(q_{l})}\\
T^{n}_2=\displaystyle{\sum_{I\in\varpi_n(3)} C_{(I)}\prod_{l=1,2,3}
 H^{i_{l}+1}\Delta_F(q_{i_l}) }\\
 \mbox{with:}\ \qquad         C_{(I)}=\displaystyle{\frac{n!}{i_1!i_2!i_3!}} \\
 T^{n}_3= \displaystyle{\sum_{J\in\varpi_{n}\atop   5\leq k\leq n}C_{ ( j_1, \dots, j_k)} \prod_{1\leq l\leq k }}
 H^{j_{l}+1} \Delta_F( q_{j_l})   
\end{array}
\label{Ti}
  \end{equation}
 In  the following proofs we rename the above decomposition (together with the equivalent one  previously given by the formula \ref{classicalDec}) as the  ``\emph{classical connected parts decomposition}''.

Moreover,  for every term in \ref{classicalDec} and \ref{Ti} (resp.  for every partition $J\in\varpi_{  n}$ i.e. every term in the sum $\displaystyle{\sum_{1\leq M\leq {n},1\leq N\leq {n}\atop  M+N\leq n+1}}$ of\ \ \ref{2.9}), we also simplify the notation   and write: 
 \begin{equation}\begin{array}{l}
\langle f_{(M)},\tau^{ n+1}f_{(N)}\rangle\ \ \ 
(\mbox{and respectively:}
\langle f_{(M)},\prod_{1\leq l\leq k}H^{j_{l}+1}f_{(N)}\rangle)\\ 
\ \ \\
\mbox{so  the $(O.S.P.n)$ conditions \ref{2.4} can be written as follows:}\\
\displaystyle{\sum_{1\leq M\leq n,1\leq N\leq n\atop   M+N\leq n+1}}\langle f_{(M)},\tau^{ n+1}f_{(N)}\rangle \geq 0
\end{array}
\label{2.12}
\end{equation}
\begin{equation}
\begin{array}{l}
(\mbox{And respectively  the corresponding connected  form of \ref{2.9} $(O.S.P.n)$ \ conditions:}  \ \ \\
\displaystyle{\sum_{1\leq M\leq {n},1\leq N\leq {n}\atop   M+N\leq n+1}}\displaystyle{\sum_{J\in\varpi_{  n} } \langle f_{(M)},\prod_{1\leq l\leq k}H^{j_{l}+1}f_{(N)}\rangle)\geq 0})\\
\mbox{or in terms of the three parts decomposition \ref{Ti}:}\\
\displaystyle{\sum_{1\leq M\leq {n},1\leq N\leq {n}\atop   M+N\leq n+1}}\langle f_{(M)}, (T^{n}_1+ T^{n}_2 +T^{n}_3)  f_{(N)}\rangle\geq 0
\end{array}
\label{2.13}
\end{equation}   
\begin{remark} \label{rem.2.1}
As one can see on the  examples \ref{matrices}  the positivity $(O.S.P.n)$ conditions for every fixed $n$ during the recursive procedure  of our proof will be given in terms of the sums of only the left upper triangular matrix elements corresponding to $M+N\leq n+1$.
\end{remark}
For example let us  write the corresponding conditions to be ensured  for $n\leq 5$: 
\begin{equation}\begin{array}{l}
 \mbox{For} \ \ n=1\ \ \  \langle f_{(1)},\tau^{  2}f_{(1)}\rangle\geq 0\ \ \ \ \\
 \mbox{For} \ \ n  =3: \\
\ \ \ \ \ \langle f_{(1)},\tau^{  2}f_{(1)}\rangle\ +\ 2\Re \langle f_{(1)},\tau^{4} f_{(3)}\rangle + \langle f_{(2)},\tau^{4}f_{(2)}\rangle\geq 0 
\end{array}
\label{2.14}
\end{equation}
\begin{equation}\begin{array}{l}
\ \  \ \ \mbox{For} \  \ n= 5 : \\
\ \langle f_{(1)},\tau^{  2}f_{(1)}\rangle
+\ 2\Re \langle f_{(1)},\tau^{4} f_{(3)}\rangle + \langle f_{(2)},\tau^{4}f_{(2)}\rangle\\
+2\Re \langle f_{(1)},\tau^{ 6} f_{(5)}\rangle + 2\Re \langle f_{(2)},\tau^{ 6} f_{(4)}\rangle+2\Re \langle f_{(3)},\tau^{ 6} f_{(3)}\rangle\geq 0
\end{array}
\label{2.15}
\end{equation}

In Appendix \ref{App.4.1} we show the following:
\begin{lemma}\label{lemma 2.1}
The  $(O.S.P.n)$ conditions for $n\leq 5$  are verified under the ``weak'' condition $\Lambda< 1/6$.  (cf.remark \ref{Rem1.1})
\label{lemma 2.1}
\end{lemma} 
In the next section and by using  the results  of the previous lemma \ref{lemma 2.1}
 as starting point we establish recurrenty 
the  O.S.P.  conditions for every $n$ under the same condition on the coupling constant: $\Lambda < 1/6$.
\section{Verification of the $(O.S.P.n)$  conditions   by the $\Phi^4_4$  solution}
\subsection{The auxiliary Lemmas}
Before  the main result given by the theorem \ref{Th.3.1}, 
we  present the following two  useful auxiliary statements.
The first one   presents  the ``complete splitting-factorization'' properties   verified by the  bounds \   $H_{min}^{n+1}$  in terms of the $H^2$-point functions. 
Moreover an  evident  bound is established  for all $n \geq 5 $ by using the reminders of  proposition \ref{prop.5.1} and definition  \ref{def.SplitBounds}.
The proof is directly  obtained   recurrently by using the definitions 
 \ref{4.95}, \ref{4.96}, \ref{4.100}.

The second Lemma relates the  non connected Green's function:   $\tau^{n+1}$ with all the ``preceding'' non connected i.e. 
$\tau^{i+1},  (i=1, 3, \dots,   n-2)$ and it constitutes the pivot of the recurrent proof of the theorem. The proof is given in 
Appendix \ref{Ap.4.3}.

\begin{lemma}\label{3.1}

\textbf{The complete splitting} 

$\forall n  \geq 7 $  the following ``complete splitting'' properties are  verified by the  bounds 
  $ H_{min}^{n+1}$  in terms of the $H^2$-point functions.
\begin{description}
\item{i)}
\begin{equation} 
 \vert H^{n+1}_{min}\vert=\displaystyle{\prod_{m=3}^{n} \delta_{m,min}\tilde {\mathcal{T}_{m}} }\displaystyle{\prod_{l=1}^n  H^2(q_{l})\Delta_F(q_{l})} 
\label{3.16}
\end{equation}
(For the number $\tilde {\mathcal T} _{n}$ cf. remark \ref{rem.4.2})
\item{ii)}
\begin{equation}
\forall  n\geq 5\quad  \quad\delta_{n, max}< 3\Lambda n (n-1)
\label{3.17}
\end{equation}
\end{description}
\end{lemma}

\begin{lemma}\label{3.2}  
 We suppose that the following properties are valid $\forall\ \ {\bar n}\leq n-2$. then:
  \begin{enumerate}
  \item{} If\   \ $H^{n+1}>0 $ , 
   \begin{description}
 \item{a)}  \begin{equation}\begin{array}{l} 
( \tau^{ n+1}-T^{n}_1)
 =\displaystyle{\sum_{I\in\varpi_n(3)}C_{(I)}\prod_{l=1,2,3}
   \tau^{i_{l}+1}}\\
 \mbox{here:}\ \qquad        C_{(I)}=\displaystyle{\frac{n!}{i_1!_2!_3!}} \ 
  \end{array} 
\label{3.18}
\end{equation}            
\item{b)}
\begin{equation}
 \forall(q,\Lambda)\in \mathcal{E}^{4n}\times]0,\ 1/6[ , \qquad  
 \tau^{ n+1}-T^{n}_1  \geq 0  
 \label{3.19}
 \end{equation}
 \end{description} 
 
   \item{}   If\  \  $H^{n+1}\ <\ 0 $ : 
   \begin{description}
 \item{a)} \label{Lemma  3.2 2.a)}
\begin{equation}\begin{array}{l} 
\displaystyle{\frac{|C^{n+1}|}{6\Lambda}}\geq \frac{n(n-1)\bar {\mathcal T} _{n}}{2}\displaystyle{\prod_{m=3}^{(n-2)} \delta_{m,min} \bar {\mathcal{T}_{m}}}\displaystyle{\prod_{l= 1}^{n }  H^2 (q_{l})\Delta_F( q_{l})^2}\\ 
\ \ \\
T^{n}_1 +T^{n}_2\geq  \frac{n(n-1)}{2} \bar {\mathcal T} _{n} H^{n-1}_{min }\displaystyle{\prod_{l=n-1,n}  H^2 (q_{l})\Delta_F( q_{l})^2}\left\{1-\frac{2\delta_{nmax}}{n(n-1)}\right\} \\
\mbox{\emph{and}} \quad\forall(q,\Lambda)\in \mathcal{E}^{4n}\times]0,\ 1/6[ ,    
  \quad                  T^{n}_1 +T^{n}_2\geq 0
\end{array} 
\label{3.20}
\end{equation}
\item{b)}
\begin{equation}\begin{array}{l}
 T_3^{n}=\displaystyle{\sum_{I\in\varpi_n(3)}C_{(I)}\prod_{l=1,2,3} 
\tilde \tau^{i_{l}+1}(q_i)}\\
\ \ \\
\mbox{here:}\ 
  \tilde \tau^{i_{l}+1}=\tau^{i_{l}+1}-T^{i_{l}+1}_1\ \ \mbox{if} \ \  l=1\\
  \ \tilde \tau^{i_{l}+1}=\tau^{i_{l}+1}\ \  \mbox{if}\  l=2,\  3\\
  \ \ \\
 \mbox{and} \   \quad\forall(q,\Lambda)\in \mathcal{E}^{4n}\times]0,\ 1/6[ ,  \quad  T_3^{n}\geq 0 
  \quad            
  \end{array}
\label{3.21}
\end{equation}
   \end{description}  
\end{enumerate}
\end{lemma}
 
\subsection{The Main result}
\begin{axiom}\
For every $n=2k+1, k\in \N$  and for all integers $M, \ N$ \   with $ 1\leq M\leq n, 1\leq N\leq n;\   M+N\leq n+1$  the following lower positive bounds are verified  by the non connected Green's functions.
\begin{enumerate}  \item{} 
\begin{equation}\begin{array}{l}
 \mbox{if}  \ \      H^{n+1}> 0  \\
\Rightarrow \qquad\forall(q,\Lambda)\in \mathcal{E}^{4n}\times]0,\ 1/6[ \ \ \\
\ \ \\
\Re\langle f_{(M)},\tau^{ n+1}f_{(N)}\rangle  
\geq h(n,\Lambda) \Vert\ f \Vert^2 (G_1)^{n-1}\ \geq 0 \\
\mbox{where} \\
\Vert f\Vert^2 =\displaystyle{\int\vert f ^{(M)}(q ) \vert^2(H^{2}\Delta_F^2)(q_M)dq_M} \\
( G_1)^{n-1}=(-1)^{(n-1)}\left\{\displaystyle{ \sup_{(i)}\int \vert f^{(i)}(q_i)\vert (H^{2}\Delta_F^2)(q_i)dq_i }\right\}^{n-1}\ \\
\mbox{and}\\   
 h(n,\Lambda) = \frac{(n-2)(n-3)}{2} \displaystyle{\prod_{m=3}^{n } \delta_{m,min} \bar {\mathcal{T}_{m}}}\times
\left\{1-\frac{2\delta_{n-2, max}}{(n-2)(n-3)}\right\}  \geq 0
\end{array}
\label{3.22}
\end{equation}

\item{} 
\begin{equation}\begin{array}{l}
\mbox{if} \quad H^{n+1}< 0\\
\Rightarrow \qquad\forall(q,\Lambda)\in \mathcal{E}^{4n}\times]0,\ 1/6[ \ \ \\
\ \ \\
\Re\langle f_{(M)},\tau^{ n+1}f_{(N)}\rangle  
\geq  \hat h(n,\Lambda) \Vert\ f \Vert^2 (G_1)^{n-1} \\
\ \ \\
\mbox{where} \Vert\ f \Vert^2  \  \mbox{and}(G_1)^{n-1} \quad\mbox{are the same as before.}\\
\mbox{and}  : \\
 \hat h(n,\Lambda) = \displaystyle{\frac{n(n-1)\bar {\mathcal T} _{n}}{2}}\left\{\displaystyle{1-\frac{2\delta_{nmax}}{n(n-1)}}\right\}\displaystyle{\prod_{m=3}^{n-2} \delta_{m,min} \bar {\mathcal{T}_{m}}}  \geq 0
\end{array}
\label{3.23}
\end{equation}
\end{enumerate}
\label{Th.3.1}
\end{axiom} 
The proof of Theorem \ref{Th.3.1} \ is presented in Appendix \ref{App.4.2}

Finally, as a corollaray we directly obtain our main result:
\begin{axiom}\
For every $n=2k+1, k\in \N$   the $(O.S.P.n)$  conditions \ref {2.12} are verified under the  following  ``weak condition''  imposed on
the physical coupling constant:
\begin{equation}\Lambda< \frac{1}{6}
\label{3.24}
\end{equation}
\label{Th.3.2}
\end{axiom}

\section*{Acknowledgments}

The author is indebted to Ph. Blanchard for his constant interest in her work. She is grateful to V. Georgescu who has followed closely her work and provided advice, suggestions and constructive criticisms. She would like also to thank B. Grammaticos for his critical reading of the successive versions of the manuscript.

 \vfill\eject

 \vfill\eject
\section{APPENDICES} 
\subsection{The first examples}
\begin{Appendix}\ \label{App.4.1}

\begin{enumerate}
\item{} \emph{The matrix representations of $\langle f_{(M)},\tau^{4}f_{(N)}\rangle
$ and $\langle f_{(M)},\tau^{ 6}f_{(N)}\rangle$. As we noticed before (cf. remark \ref{rem.2.1}) the matrix elements denoted by the symbol (*)
are not taken into account in formulas \ref{2.12} and \ref{2.13} because they are such that $M+N> n+1$.}
\begin{equation}
 P_3=\begin{pmatrix}(1,1)&0&(1,3)\cr
\hfill&\hfill&\cr
0&(2,2)&0\cr
\hfill&\hfill&\cr
(3,1)&0&( *)\cr
\hfill&\hfill&\hfill\cr
\end{pmatrix}
\quad P_5=\begin{pmatrix}
(1,1)&0&(1,3)&0&(1,5)\cr
\hfill&\hfill&\hfill&\hfill&\cr
0&(2,2)&0&(2,4)&0\cr
\hfill&\hfill&\hfill&\hfill\cr
(3,1)&0&(3,3)&0&(*)\cr
\hfill&\hfill&\hfill&\hfill&\cr
0&(4,2)&0&(*)&0\cr
\hfill&\hfill&\hfill&\hfill&\cr
(5,1)&0&(*)&0&(*)\cr
\end{pmatrix}
\label{matrices}
\end{equation}
\item{}
\textbf{Proof of Lemma }\ref{lemma 2.1}
 (The proof for $n\leq 5$ )
 \begin{description}
\item{a)}\emph{ For $n=1$ by using the positivity  of $H^2$ we have trivially: }
\begin{equation}
\langle f_{(1)},\tau^{  2}f_{(1)}\rangle=\langle f_{(1)}H^{  2}f_{(1)}\rangle =\int\vert f^{(1)}(q_1)\vert^2H^{  2}(q_1)[\Delta_F(q_1)]^2dq_1\geq 0
\label{4.26}
\end{equation}
\item{b)}
\emph{For $n=3$ we estimate every connected contribution, of}
\begin{equation}
 2\Re \langle f_{(1)},\tau^{4} f_{(3)}\rangle \ \mbox{\emph{and}} \ \langle f_{(2)},\tau^{4}f_{(2)}\rangle 
\label{4.27}
\end{equation}
\emph{By using    the decomposition of $\tau^{4}$  into  its connected parts: $$H^4\prod_{l=1}^3\Delta_{F}(q_l)\ \  \mbox{\emph{and}} \prod_{l=1,2,3} (H^2\prod_{l}^3\Delta_{F}^2)(q_l) $$ we have}
\begin{equation}
 \Re \langle f_{(1)},\tau^{4}f_{3)}\rangle =
 \Re \langle f_{(1)},H^{4}\prod_{l=1}^3 \Delta_{F_{(l)}} f_{(3)}\rangle +
 \Re \langle  f_{(1)},\prod_{l=1}^3  H^2_{(l)}\Delta_{F}^2 f_{(3)}\rangle 
\label{tau4}
\end{equation}
\emph{We consider the first term of the r.h.s. of \ref{tau4}. By application of splitting and sign properties of $H^4$ (cf.definition \ref{def.4.2})
  together with the factorized test functions (in view of the nuclear theorem as explained in definition \ref{def.factor} )
   and by Fubini's theorem we write:}
\begin{equation}
\begin{array}{l}
 \Re \langle  f_{(1)} ,H^{4}  f_{(3)}\rangle =\displaystyle{ - \Re \int {\bar f^{(1)}(q_1)}\delta_{3}(q_{(3)}) \prod_{l=1}^3  (H^{2}\Delta_F^2)(q_l){f^{(l)}}(q_l)dq_l}\\
\geq-|\Re\displaystyle{\int}\displaystyle{ \prod_{l=2}^{3 } (f^{(l)}H^{2}\Delta_F^2)(q_l)dq_{l}\left\{\int\vert f^{(1)} (q_{1})
 \vert^2\delta_{3}(q)(H^{2}\Delta_F^2)(q_1)dq_{1}\right\}}|
\end{array}
\label{4.29}
\end{equation}
 \emph{The integral with respact to $q_1$ is positive and taking into account definition \ref{def.SplitBounds} for the upper bound of  the splitting function:}\begin{equation}\begin{array}{l}
\forall(q,\Lambda)\in \mathcal{E}^{12}\times]0, 0.04]  \\
	\delta_{3}(q,\Lambda)\leq \delta_{3,max}<6\Lambda
	\end{array}
	 \label{4.30}
 \end{equation} \emph{we finally obtain:}
\begin{equation} 
\Re \langle f_{(1)},H^{4} f_{(3)}\rangle\geq -  6\Lambda\ \Vert f^{(1)}\Vert^2 (G_1)^2 
 \label{4.31}
\end{equation}
\emph{where we used the following notations:}
\begin{equation} 
\Vert f \Vert^2 =\displaystyle{\int\vert f^{(1)} (q_{1}) \vert^2(H^{2}\Delta_F^2)(q_1)dq_{1}} 
\label{4.32}        
\end{equation}
\emph{and},
\begin{equation}
( G_1)^{2}=(-1)^{2}\left\{\displaystyle{ \sup_{(i)}\int \vert f^{(i)}(q_i)\vert (H^{2}\Delta_F^2)(q_i)dq_i }\right\}^{2}\ \ \ \  \ 
 \label{4.33}
\end{equation} 
\emph{In an analogous way, and by using the positive sign of $H^2$ point function the second term of the r.h.s. of \ref{tau4} yields:}
 
\begin{equation}
\begin{array}{l}
\Re \langle f_{(1)},\prod_{l=1}^3  H^2_{(l)}\Delta_{F_{(l)}}^2 f_{(3)}\rangle \\
\geq  \displaystyle{\Re\int\vert f^{(1)} (q_{1}) \vert^2H^{ 2}\Delta_F^2dq_{1}} 
(-1)^{2}\left\{\displaystyle{ \sup_{(l)}\int \vert f^{(l)}(q_l)\vert (H^{2}\Delta_F^2)(q_l)dq_l }\right\}^{2}\\ 
\geq  \Vert f\Vert^2(G_1)^2\geq 0
\end{array}
\label{4.34}
\end{equation}
\emph{Now, by inserting the results \ref{4.31}  and \ref{4.34} in  \ref{tau4}
we obtain:}
\begin{equation}\begin{array}{l}
 \Re \langle f_{(1)},\tau^{4}f_{(3)}\rangle \geq {(1-6\Lambda)}\Vert f \Vert^2(G_1)^2\geq 0 \\
 \mbox{\emph{under the condition}} \ \ \ \Lambda<\frac{1}{6}
 \end{array}
 \label{4.35}
\end{equation}
\emph{By using analogous arguments to the ones we previously presented, a similar result is obtained for the contribution of the term $\langle f_{(2)},\tau^{4}f_{(2)}\rangle\geq 0 $.
  Precisely:}
\begin{equation}
\langle f_{(2)},\tau^{4}f_{(2)}\rangle\geq (1-6\Lambda)\Vert f \Vert^2(G_1)^2\geq 0
 \label{4.36}
\end{equation}
\emph{under the condition   \ \ $ \Lambda<\frac{1}{6}$.}

\emph{ Finally:}
\begin{equation}\begin{array}{l}
\displaystyle{\sum_{1\leq M\leq {3},1\leq N\leq {3}\atop   M+N\leq 4}}\displaystyle{\sum_{\varpi_{n}}\langle f_{(N)},\tau^{4}f_{(M)}\rangle} \geq 0\ \\
\mbox{\emph{under the condition}} \ \ \ \Lambda<\frac{1}{6}\\
\end{array}
 \label{compltau4}
\end{equation}

\item{c)} \emph{ For $n= 5$}
 
 \emph{Following  
\ref{2.15} the supplementary condition to ensure is the positivity of the sum:}
\begin{equation}
 2\Re \langle f_{(1)},\tau^{ 6} f_{(5)}\rangle + 2\Re \langle f_{(2)},\tau^{ 6}f_{(4)}\rangle+\langle f_{(3)},\tau^{ 6}f_{(3)}\rangle\geq 0
 \label{4.38}
 \end{equation}
\emph{with:}
 \begin{equation}\begin{array}{l}
  \tau^{ 6}=H^6 \prod_{l=1}^5 \Delta_{F}(q_l)\\
 \hspace{2cm}+\frac{5!}{3!2}H^4\prod_{i=1}^3 \Delta_{F}(q_i)\prod_{l=4}^5  (H^2  \Delta_{F}^2)(q_l) \\
\hspace{4cm} +\frac{5!}{3!2}\prod_{i=1}^3 (H^2 \Delta_{F}^2)(q_i)\prod_{l=4}^5  (H^2  \Delta_{F}^2)(q_l) 
 \end{array}
 \label{4.39}
\end{equation}
\emph{The procedure being similar for each one of the terms in 
\ref{4.38}
 we give the proof only for $2\Re \langle f_{(1)},\tau^{ 6} f_{(5)}\rangle$, and for each term of the connected parts decomposition \ref{4.39}}.
\begin{itemize}
\item{} \emph{By using the splitting and sign properties of $H^6$ and $H^4$ (cf. def.\ref{def.4.2}c) we have:}
\begin{equation}
H^6=\delta_5(q_{(5)})\delta_3(q_{(3)})\prod_{l=1}^5  (H^2  \Delta_{F}^2)(q_l)  
 \label{H6}
\end{equation}
\emph{Then, as before we apply the factorization of the test functions and Fubini's theorem on the first term connected part contribution in \ref{4.39}}
 \emph{and write as follows:}
 \begin{equation}
\begin{array}{l}
 \Re \langle f_{(1)},H^6\prod \Delta_{F} f_{(5)}\rangle =\Re\displaystyle{\int}\displaystyle{ \prod_{l=2}^{5 } f^{(1)}(q_l)}\\
\displaystyle\left\{\int\vert f^{(1)} (q_{1})
 \vert^2\delta_5(q_{(5)})\delta_3(q_{(3)})(H^{2}\Delta_F^2)(q_1)dq_{1}\right\} 
 (H^{2}\Delta_F^2)(q_l)dq_{l}\\
 \end{array}
\label{4.41}
\end{equation}
\emph{The integral with respact to $q_1$ being positive and taking into account definition \ref{def.SplitBounds} for the lower bounds of  the (solution in \cite{MM7}) splitting functions, precisely: }\begin{equation}\begin{array}{l}
\forall(q,\Lambda)\in\mathcal{E}^{12}\times]0,0.04] 
	\ \ \quad\delta_{3}(q,\Lambda)\geq \delta_{3,min}(\Lambda)\\
	\mbox{and}\ \ \forall(q,\Lambda)\in\mathcal{E}^{20}\times]0,0.04] \ \quad  \delta_{5}(q,\Lambda)\geq \delta_{5,min}(\Lambda)
	\end{array} 
	\label{4.42}
 \end{equation} \emph{ the factorization of the test functions and Fubini theorem, we finally obtain:}
\begin{equation}
\begin{array}{l}
 \Re \langle f_{(1)},H^{ 6}\prod \Delta_{F } f_{(5)}\rangle \geq    \delta_{5, min}\delta_{3, min}(\Lambda)\Vert f \Vert^2(G_1)^4 \geq 0 
\end{array}
\label{Positivity.6}
\end{equation}
\emph{without any supplementary condition on $\Lambda$.}

\emph{We also notice that the real positive numbers  $\Vert f \Vert^2$ (and resp. $(G_1)^4$) are defined by analogy with \ref{4.32} (and \ref{4.33} respectively)} 
\item{} \emph{For  the second  and third terms we proceed by analogy. We take the sum of them and by using the sign-splitting property of $H^4$ and the corresponding results of \ref{4.31} \ref{4.34} and \ref{4.35} we write:}
\begin{equation}\begin{array}{l}
\frac{5!}{3!2}\Re \langle f_{(1)}\left\{-|H^4| +\prod H^2 \Delta_{F }\right\}\prod  \Delta_{F}f_{(5)}\rangle\\
\times\displaystyle{\prod_{l=4}^5\int f^{(l)}(q_l) (H^{2}\Delta_F^2)(q_l)dq_{l}}\\ 
= \frac{5!}{3!2}\Re \langle f_{(1)},\tau^{4}f_{(3)}\rangle \times\displaystyle{\prod_{l=4}^5\int \bar f^{(l)}(q_l) (H^{2}\Delta_F^2)(q_l)dq_{l}}\\
\geq \frac{5!}{3!2}{(1-6\Lambda)}\Vert f\Vert^2(G_1)^4\geq 0\hspace{0.21cm} \mbox{\emph{under the condition}} \ \ \ 0<\Lambda<\frac{1}{6}\\
\emph{ \mbox{Finally}}\\
 2\Re \langle f_{(1)},\tau^{ 6} f_{(5)}\rangle \ \geq \\
  \geq     \left\{\delta_{5, min}\delta_{3, min}(\Lambda)  +\frac{5!}{3!2 }{(1-6\Lambda)}\right\}\Vert f\Vert^2(G_1)^4\geq 0\\ 
  \mbox{\emph{under the condition}} \ \ \ 0< \Lambda <\frac{1}{6}
 \end{array} 
 \label{4.44}
 \end{equation}
 \end{itemize}
\end{description}

\emph{\textbf{Conclusion}}

\emph{From the results of a) b) c) we finally obtain:}
 \begin{equation}\begin{array}{l}
\displaystyle{\sum_{1\leq M\leq {5},1\leq N\leq {5}\atop   M+N\leq 6}}\displaystyle{\sum_{\varpi_{n}}\langle f_{(N)},\tau^{ 6}f_{(M)}\rangle} \geq 0\ \\
\mbox{\emph{under the condition}} \ \ \ \Lambda<\frac{1}{6}\\
 \end{array} 
 \label{4.45}
 \end{equation} 
$\hspace{10cm} \ \blacksquare $
  \end{enumerate}

\end{Appendix}

\subsection{The proof of Theorem \ref{Th.3.1}}

\begin{Appendix}\ \label{App.4.2}

\emph{ We suppose that the statement holds  $\forall \ \bar n\leq n-2$ }
\begin{enumerate}
\item{}
\begin{description}
\item{a)} 
\emph{ Let \ \ $H^{n+1}>0$ \  \ ( or  $T^{n}_1>0$)}.

\emph{For an  arbitrary couple $ (M, N)$ with   $M\leq n,  N\leq n$ and $N+M=n+1$
 we consider the corresponding first  term in the sum $\sum_{\varpi_{n}}$
of equation \ref{classicalDec} (that means when $k=1$).}
\emph{We  suppose that the momentum variables are ordered and the test fuctions factorized then by Fubini's theorem we write:}
 \begin{equation} \begin{array}{l}
\Re\langle f_{(M)} H^{n+1}\prod\Delta_Ff_{(N)} \rangle\\
=\Re \displaystyle{\int\left\{I_{nt} (q_{(n-1)})\right\}\prod_{i=1}^{M-1}\overline{f^{(i)}(q_i)}\Delta_F(q_i)dq_i\prod_{l=2 }^{N}}f^{(l)}(q_l)\Delta_F(q_l)dq_l\\ 
\mbox{\emph{Here:}}\\ 
I_{nt} (q_{(n-1)})= \displaystyle{\int\vert f^{(M)}(q_M)\vert^2 H^{(n+1)}(q_{(n)})\Delta_F( q_M)dq_M }\\
\end{array}
\label{4.46}
\end{equation}    
 \emph{The positivity of the  integrand of $I_{nt} (q_{(n-1)})$  allows us  to take the lower bound of $ H^{n+1}$ 
by using lemma \ref{3.1} (of complete splitting) and by following an analogous procedure as the one of $H^6$ we obtain:}
\begin{equation}\begin{array}{l}
 \Re\langle f_{(M)}   H^{n+1}\displaystyle{\prod \Delta_Ff_{(N)}  \rangle}\\
 \geq \displaystyle{\prod_{m=3}^{n} \delta_{m,min} \bar {\mathcal{T}_{m}}} \Vert\ f \Vert^2 
  \displaystyle{\prod_{i=1}^{M-1}(-1)^{(i)}\sup_{i}\int\vert{f^{(i)}(q_i)}\vert H^2(q_{i})[\Delta_F(q_{i})]^2dq_i}\\
 \times \displaystyle{\prod_{l=2 }^{N}(-1)^{(l)}\vert f^{(l)}(q_l)\vert H^2(q_{l})[\Delta_F(q_{l})]^2dq_l}\\ 
\geq
\displaystyle{\prod_{m=3}^{n} \delta_{m,min} \bar {\mathcal{T}_{m}}} \Vert\ f \Vert^2  (-1)^{(n-1)}\left\{\displaystyle{ \sup_{(i)}\int \vert f^{(i)}(q_i)\vert (H^{2}\Delta_F^2)(q_i)dq_i }\right\}^{n-1}\\  \geq \displaystyle{\prod_{m=3}^{n} \delta_{m,min} \bar {\mathcal{T}_{m}}} \Vert\ f \Vert^2 (G_1)^{n-1}\geq 0
\end{array}
\label{4.47}
\end{equation}
\emph{So we obtain the positivity without any supplementary condition on the coupling constant than $0<\Lambda < \frac{1}{6}$ which is required by the recurrence hypothesis.}
\emph{We notice that $\Vert f_1\Vert^2$ (and resp. $(G_1)^{(n-1}$) are always defined by analogy with \ref{4.32} (and \ref{4.33} respectively).}

$\hspace{10cm} \ \blacksquare $
\item{b)} 
\emph{Taking into account the notations \ref{Ti} we have to show that:}
 \begin{equation}\begin{array}{l} 
 \forall  (M,\ N)\  \mbox{\emph{such that}}\ \  1\ \leq M\ \leq \ \frac{n+1}{2}\ \leq N\leq n,\\
  \ \ \forall (q,\Lambda)\in \mathcal{E}^{4n}\times]0,\ 1/6[ , \ \\
\displaystyle{\sum_{1\leq M\leq {n},1\leq N\leq {n}\atop   N+M=n+1}}\langle f_{(M)}, ( T^{n}_2 +T^{n}_3) f_{(N)}\rangle\geq 0 
\end{array}
\label{4.48}
\end{equation}
\emph{or equivalently by application of Lemma \ref{3.2} 1 a)  equation  
\ref{3.18}  show that:}
\begin{equation}\begin{array}{l}
\forall (q,\Lambda)\in \mathcal{E}^{4n}\times]0,\ 1/6[ \\
\Re\langle f_{(M)}\displaystyle{\sum_{I\in\varpi_n(3)}C_{(I)}}\prod_{l=1}^{3}
 \tau^{i_{l}+1}f_{(N)}\rangle\geq 0\\
 \end{array}
\label{4.49}
\end{equation} 
\emph{We always suppose that the moments are ordered and the test fuctions factorized, so by Fubini's theorem we write:}
 \begin{equation} \begin{array}{l}
\Re\langle f_{(M)}\displaystyle{\sum_{I\in\varpi_n(3)}C_{(I)}}\prod_{l=1}^{3}
 \tau^{i_{l}+1}f_{(N)}\rangle\\
 =\Re\displaystyle{\int\left\{I_{nt} (q_{(n-1)})\right\}\prod_{i=1}^{M-1}\overline{f^{(i)}(q_i)}dq_i\prod_{l=2 }^{N}}{f^{(l)}(q_l)}dq_l\\
\end{array}
\label{4.50}
\end{equation}
\emph{Where:} 
\begin{equation} 
I_{nt} (q_{(n-1)})=\int\vert f^{M)}(q_M)\vert^2 \displaystyle{\sum_{I\in\varpi_n(3)}C_{(I)}\prod_{l=1,2,3}
   \tau^{i_{l}+1}(q_{(i_l)})dq_M}  
\label{4.51}
\end{equation}
\emph{Notice that following   Lemma \ref{3.2} 1 b)  (and the recurrence hypothesis)  each one of the terms  and consequently the sum itself are positive}.
\emph{So a lower bound of the sum could be  the ``first'' term-contribution}
 $\bar I=(n-2, 1, 1)$ \emph{to obtain:} 
\begin{equation}\begin{array}{l}
I_{nt} (q_{(n-1)})\geq C_{\bar I=(n-2, 1, 1)}\displaystyle{\int\vert f^{(M)}(q_M)\vert^2
   \tau^{n-1}(q_{(n-2)})dq_M}\\
\qquad\qquad\qquad \times\displaystyle{\prod_{l=n-1}^{n}H^2(q_l)\Delta_{F}^2(q_l)}
\end{array}
\label{4.52}
\end{equation}
\emph{Now by \ref{Ti}
 we have:}
\begin{equation}
 \tau^{(n-1)}=T^{(n-2)}_1+T^{(n-2)}_2+T^{(n-2)}_3 
 \label{4.53}
\end{equation}
\emph{But, $H^{n-1}<0$ so by the recurrence hypothesis of  lemma \ref{3.2} 2.a and b)} \emph{(precisely on $\delta_{n-2}$)}, \emph{and then by application of Lemma \ref{3.1} on $H^{(n-3)}>0$, we obtain:}
 \begin{equation}\begin{array}{l}
\forall(q,\Lambda)\in \mathcal{E}^{4n}\times]0,\ 1/6[  \\
T^{(n-2)}_1 +T^{(n-2)}_2\geq  \frac{(n-2)(n-3)}{2} \bar {\mathcal T} _{n-2}\displaystyle{\prod_{m=3}^{n-4} \delta_{m,min} \bar {\mathcal{T}_{m}}}\\
\times\displaystyle{\prod_{l=1}^{n-2}  H^2 (q_{l})[\Delta_F( q_{l})]^2}\left\{1-\frac{2\delta_{n-2, max}}{(n-2)(n-3)}\right\}  \geq 0\\
 \qquad \qquad \mbox{\emph{and}}  \quad\ T^{(n-2)}_3\geq 0\\
 \end{array}
 \label{4.54}
\end{equation}

\emph{So, by taking into account   \ref{4.54} inside \ref{4.53},  a lower bound of $I_{nt} (q_{(n-1)})$ is obtained that we insert in equation \ref{4.51}. Finally, we add the proof  of a)  \ref{4.47} for $T^{n}_1$, and the result is as follows:}
\begin{equation}\begin{array}{l} 
  \forall \ \ \mbox{\emph{positive integers}}\  N, M,\  \mbox{\emph{such that}} \ 1\leq M\leq {n},1\leq N\leq {n}\\
   \mbox{\emph{with:}}\   N+M=n+1  \ \ \mbox{\emph{and}}
  \  \forall (q,\Lambda)\in \mathcal{E}^{4n}\times]0,\ 1/6[ \\
  \ \ \\
\Re \langle f_{(M)}, (T^{n}_1+ T^{n}_2 +T^{n}_3) f_{(N)}\rangle\geq  h(n,\Lambda) \Vert\ f \Vert^2 (G_1)^{n-1}\ \geq 0 \\
\ \ \\
\mbox{\emph{where:}}\\
 h(n,\Lambda) = \frac{(n-2)(n-3)}{2} \displaystyle{\prod_{m=3}^{n } \delta_{m,min} \bar {\mathcal{T}_{m}}}\times
\left\{1-\frac{2\delta_{n-2, max}}{(n-2)(n-3)}\right\}  \geq 0
\end{array}
\label{4.55}
\end{equation}
\emph{This completes the  proof of the theorem \ref{Th.3.1} for $H^{n+1}>0$.}

$\hspace{10cm}\ \blacksquare$ 
\end{description}

\item{}  \emph{Case $H^{n+1}<0$. }
\begin{description}
 \item{a)}\emph{ We first have to show the positivity of:}
$$2\Re\langle f_{(M)} , (T^{n}_1+ T^{n}_2 )  f_{(N)}  \rangle $$
\emph{We  suppose always that the moments are ordered and the test fuctions factorized then by Fubini's theorem we write:}
 \begin{equation} \begin{array}{l}
\Re\langle f_{(M)} (T^{n}_1+ T^{n}_2 ) f_{(N)} \rangle=\\
=\Re \displaystyle{\int}
\left\{I_{nt} (q_{(n-1)})\right\}\displaystyle{\prod_{i=1}^{M-1}\overline{f^{(i)}(q_i)} dq_{i}\prod_{l=2 }^{N}}{f^{(l)}(q_l)}dq_l\\
\mbox{\emph{Here:}}\\
I_{nt} (q_{(n-1)})= \displaystyle{\int\vert f^{(M)}(q_M)\vert^2 (T^{n}_1+ T^{n}_2 )
(q_{(n)})dq_M }\\
\end{array}
\label{4.56}
\end{equation}   
  \emph{By using  Lemma \ref{3.2} 2.a)  we apply  the positive lower bound (cf.\ref{3.20})   of $T^{n}_1+ T^{n}_2$, and then by application of Lemma 
  \ref{3.1} on $H^{n-1}$, we  finally obtain :}
 \begin{equation}\begin{array}{l}
\Re\langle f_{(M)} (T^{n}_1+ T^{n}_2 ) f_{(N)} \rangle \\
\geq\Vert\ f \Vert^2  (-1)^{(n-1)}\left\{\displaystyle{ \sup_{(i)}\int \vert f^{(i)}(q_i)\vert (H^{2}\Delta_F^2)(q_i)dq_i }\right\}^{n-1}\\
\times \displaystyle{\frac{n(n-1)\bar {\mathcal T} _{n}}{2}}\left\{\displaystyle{1-\frac{2\delta_{nmax}}{n(n-1)}}\right\}\displaystyle{\prod_{m=3}^{n-2} \delta_{m,min} \bar {\mathcal{T}_{m}}}  \geq 0\\
\mbox{\emph{under the condition }} \quad  0<\Lambda < \frac{1} {6}
\end{array}
\label{4.57}
\end{equation}
$\hspace{10cm} \ \blacksquare $

 \item{b)} \emph{The   third term $T^{n}_3$ in \ref{2.13} of the remaining connected parts is exactly the sum appearing on the r.h.s. of \ref{3.21}.
 By the recurrence hypothesis every term of this sum (and evidently the sum itself) is positive (Lemma \ref{3.2} 2 b)  (for $\Lambda<1/6$) so we can take a lower bound of this sum by the ``first term'' $\tilde \tau^{n-1}=\tau^{n-1}-T^{n-1}_1$  and the recurrence hypothesis  of the positive $H^{n-1}$ by proceeding as in the  previous proof of $H^{n+1}>0$). The remaining procedure is then analogous to that of  $T^{n }_2 +T^{n }_3$ 
(cf. results: \ref{4.47} \ref{4.48}}):
\begin{equation} \begin{array}{l}
\langle f_{(M)} \displaystyle{\sum_{I\in\varpi_n(3)}C_{ (I)}}\prod_{l=1}^{3}
\tilde \tau^{i_{l}+1}f_N\rangle \\
\geq \langle f_{(M)}, ( T^{n-2}_2 +T^{n-2}_3) f_{(N)}\rangle\geq 0
\qquad \ \forall  \Lambda \in \ ]0,  \frac{1}{6}[
 \end{array}
 \label{4.58}
 \end{equation}
\end{description}
$\hspace{10cm} \ \blacksquare $
\end{enumerate}
\end{Appendix}

\subsection{Proof of Lemma \ref{3.2}}
\begin{Appendix} \label{Ap.4.3}

\emph{We suppose that the Lemmas \ref{3.1}, \ref{3.2} are verified $\forall\  {\bar n}\leq n-2$. Then, }
 \begin{enumerate}
\item{}  
 if\   $H^{n+1}>0 $ ,   \emph{ we show that:}
  \begin{description}
 \item{a)}  \begin{equation}\begin{array}{l} 
( \tau^{ n+1}-T^{n}_1)
 =\displaystyle{\sum_{I\in\varpi_n(3)}C_{(I)}\prod_{l=1,2,3}
  \tau^{i_{l}+1}}\\
 \mbox{with:}\ \qquad        C_{(I)}=\displaystyle{\frac{n!}{i_1!_2!_3!}} \\ 
\mbox{\emph{or equivalently, following formulas \ref{Ti}}}\\
T^{n}_2 +T^{n}_3=\displaystyle{\sum_{I\in\varpi_n(3)} C_{(I)}\prod_{l=1,2,3}
  \tau^{i_{l}+1}}
\end{array} 
\label{4.59}
\end{equation} 
\emph{In other words we have to reformulate the ``classical' decomposition of every non connected Green's function $\tau^{n+1}$ of definition \ref{classicalDec} to  a ``tree'' 
type   (cf. \ref{tree.dec.}) recursive expression in terms of the  preceding non connected $\tau^{i+1}$'s\ \  $\forall  i\leq n-2 $}.
\begin{remark}
\emph{We recall that in the standard definition of \ref{tree.dec.} (cf.\cite{MM1} c)}
 \begin{equation}\begin{array}{l}
\forall I\in \varpi_{n}(3) \  \mbox{\emph{with}}\  I= (I_1,   I_2,  I_3), { \rm Card}{I}_l =i_l \ \mbox{\emph{where}} \   i_l, \  l \in (1, 2, 3)  \\
\mbox{\emph{are odd integers such that:}} \ 
  i_1\geq i_2\geq i_3, \mbox{\emph{and}}\ \displaystyle{\sum_{l=1}^{3}i_l=n.}  
  \end{array}
 \label{4.60} 
 \end{equation}
\end{remark}

\emph{The starting point is the tree structure of $T^{n}_2$ namely:}
\begin{equation}
T^{n}_2=\displaystyle{\sum_{I\in\varpi_n(3)}C_{(I)}\prod_{l=1,2,3}H^{i_l+1}\Delta_F(q_{i_l})}= \frac{C^{n+1}}{-6\Lambda}
\label{T{n}_2}
\end{equation} 
\emph{This expression translates the first step decomposition of $\tau^{n+1}$ as a sum of three - factors' products (or ``tree type in smaller non connected parts'').}

\emph{Before giving the details of the proof let us introduce some useful notations  in order to display the relationship betheen the corresponding terms of $T^{n}_2$ and $T^{n}_3$}:

\begin{description}
\item{ i)} 
\emph{Notation for every term  of $T^{n}_2$ :} 
\begin{equation}
T^{n, (i_1,1)}_{3,3}(q_{(n)})=C_{(I)}\displaystyle{\prod_{l=1,2,3}H^{i_l+1}\Delta_F(q_{i_l})}
\label{T^{n, (i_1,1)}_{3,3}}
\end{equation}
\emph{In other words we shall often describe $T^{n}_2$ as follows:}
\begin{equation}
T^{n}_2=\displaystyle{\sum_{ I\in\varpi_n(3)}C_{(I)}T^{n, (  i_1,1)}_{3,3}}
\label{T{n}_2 bis}
\end{equation} 
\begin{equation}\begin{array}{l}
\mbox{\emph{Example: If $ i_1=n-2$, then}}\\
T^{n, (n-2,1)}_{3,3}(q_{(n)})=\displaystyle{\frac{n!}{(n-2)!2}}H^{n-1}(q_{(n-2)})\displaystyle{\prod_{l=n-1}^{n}H^2(q_{l})\Delta_F(q_{l})^2}
\end{array}
\label{1rstterm} 
\end{equation}
\item{ii) }
$ \forall \ \mbox{\emph{fixed}}\  k=3, 5, 7, \dots i_1, i_1+2, $
\emph{we use the following notation of the $(k-2)^{th}$ order step of the classical type development of $H^{i_1+1}$ (or of the decomposition \ref{Ti} of $\tau^{i_1+1}$) in terms of sums of ``tree type'' products. It corresponds to the $k^{th}$ order ``new tree type decomposition of $\tau^{n+1}$ :} 
 \begin{equation} \begin{array}{l}
T^{n, (i_1,k-2)}_{3,k}= \displaystyle{\frac{n!}{i_1!i_2!i_3! }}  
C_{ (\hat  j_1, \dots, \hat  j_k)} \prod_{1\leq l\leq k}H^{(\hat j_{l}+1)}\\
\mbox{here:} \ \ C_{ (\hat  j_1, \dots, \hat  j_k)} = \displaystyle{\frac{i_1!}{\hat {j}_1!\dots \hat {j}_{k-1}!, \hat {j}_k!} }
\end{array}
\label{4.66}
\end{equation}
\end{description}
\emph{Then, starting from the first term-triplet of the sum \ref{T{n}_2} precisely: $T^{n, (n-2, 1)}_{3,3}$ (given previously by \ref{1rstterm})  
we proceed  in a decreasing order of $i_1$'s and repeat for  every triplet  in formula of \ref{T{n}_2} the same recurrent procedure which follows:}

\emph{By keeping invariant the product $H^{i_2+1}H^{i_3+1}$ we obtain the   contribution of the triplet to the step $k=5$ (reminder: k=5 factors ) of the classical expansion of  $\tau^{n+1}$, by writing the tree type expansion $T^{i^{(1)}_1}_2$, of $H^{i_1 +1}$, so:}
\begin{equation}
T^{n, (i_1,3)}_{3,5}= \displaystyle{\frac{n!\prod_{l=2,3}H^{i_l+1}}{i_1!_2!_3! }}  
\displaystyle{\sum_{I\in\varpi_{{i}_{1}(3)}}\displaystyle{\frac{i_1!}{i^{(1)}_1!i^{(1)}_2!i^{(1)}_3!}}\prod_{l=1,2,3}H^{i^{(1)}_l+1}\Delta_F(q_{i_l})}
\label{4.67}
\end{equation}
\emph{For the following step $k=7$ we apply by decreasing order of every triplet the same tree type expansion of every
first factor $H^{i^{(1)}_l+1} $, and so on $\dots$} 
\emph{In other words we  develop successively the $H^{i_1+1}$  (the first factor) step by step in ``tree type'' $T^{i_1}_2$ or $C^{i_1+1}/(-6\Lambda)$s  sums of products of ``smaller'' and ``smaller'' connected Green's functions. Every new step $k$ is obtained from the step $k-2$ by developing the first $H^{i^{(k-4)}_l+1} $'s by decreasing order of the  triplets of the previous $T^{i_1^{(k-4)}}_2$.}
\emph{So, we obtain in a precise order the terms: $T^{n, (i_1,k-2)}_{3,k}H^{i_2+1}H^{i_3+1}$ from $5\leq k$ up to $k\leq  i_1+2$  (cf.definitions \ref{Ti}). }

\emph{The sum, of all the intermediate steps at this level  yields:}
\begin{equation}\big (\displaystyle{\sum_{k=3, 5, \dots, i_1+2}T^{n, (i_1,k-2)}_{3,k}}\big )H^{i_2+1}H^{i_3+1}=\tau^{i_1+1}H^{i_2+1}H^{i_3+1}\\
\label{4.68}
 \end{equation}
 \emph{By keeping unchanged $\tau^{i_1+1}H^{i_3+1}$ we continue in a analogous way  the reconstruction of $\tau^{i_2+1}$
 by using the ``new type in smaller connected parts decomposition'' successively of $H^{i_2+1}$ in terms of sums of tree type products. The result is equal to $\tau^{i_1+1}\tau^{i_2+1}H^{i_3+1 }$. The proof of every triplet  reconstruction is completed by the synthesis in an analogous way of $   \tau^{i_3+1}$}.    
   
   \hspace{10cm}$\blacksquare$
  \end{description}                                                             

 b) \emph{We take into account the proven property \ref{4.59}
 and the recurrence hypothesis of both lemmas for every $i_1\leq n-2$
 then the proof of positivity under the condition $\Lambda\leq 1/6$ is automatically obtained:}
\begin{equation}
 \forall(q,\Lambda)\in \mathcal{E}^{4n}\times]0,\ 1/6[ , \qquad  
 \tau^{ n+1}-T^{n}_1  \geq 0    
\label{4.69}
 \end{equation}
                                                                \hspace{10cm}$\blacksquare$
 



\item{} 
 Let\   $H^{n+1}<0 $. \ 
 
 a)\emph{ We first show the properties \ref{3.20}}.  

The lower bound of the tree global term $\frac{|C^{n+1}|}{6\Lambda}$
\emph{We consider the definition \ref{Ti}  of $T^{n}_1$ and $T^{n}_2$ . By using on one hand the definitions \ref{def.4.2}, \ref{splitting} of  the splitting of $H^{n+1}$ and on the other hand the definition \ref{tree.dec.} 
of the tree term of the mapping $C^{n+1}$, we associate with every connected contribution of the tree $C^{n+1}$, the corresponding connected contribution $\displaystyle{\prod_{1\leq l\leq k}H^{j_{l}+1}}$ in the sum $\sum_{I\in\varpi_{n}(3)}$,  
 and obtain:}
\begin{equation}\begin{array}{l}
 T^{n}_1+  T^{n}_2=H^{n+1}\displaystyle{\prod_{l=1}^n[\Delta_F(  q_{l}]}+\displaystyle{\sum_{I\in\varpi_n(3)}C_{ (I)}\prod_{l=1,2,3}
 H^{i_{l}+1}\Delta_F(  q_{i_l})}\\
=  \displaystyle{\frac{|C^{n+1}|}{6\Lambda}\left\{1-2\frac{\delta_{n}(q,\Lambda)}
	{ n(n-1)}\right\}}
\end{array}
\label{4.69}
\end{equation}
\emph{In order to obtain the positivity property of \ref{3.20}, we take the upper bound of the splitting function (cf. \ref{splitting} in the reminders):}
\begin{equation}\delta_{n,min}(\Lambda) \leq \delta_{n}(\tilde q, \Lambda)\leq
	\delta_{n,max} (\Lambda)\end{equation}       
\emph{then, by application of the bound lemma \ref{3.1} ii) we verify:}	
\begin{equation}\begin{array}{l}
T^{n}_1+  T^{n}_2 \geq 0\\
\mbox{\emph{under the condition}}\ \ \ \Lambda< 1/6
\end{array}
 \label{4.71}
\end{equation}
   \hspace{10cm}$\blacksquare$

\emph{b)} 
\emph{The ``tree decomposition'' of the term  $T_3^{n}$ (cf.\ref{Ti})
 in \ref{3.21}  is established by the same procedure  
 developed before in the case of the corresponding
$T_{3}^{n}$ of   $H^{n+1}>0$, with a small  difference:
the analogous triplets do not contain their first factor: 
\begin{equation} 
T^{i_1}_1=H^{i_1+1}\displaystyle{\prod_{j=1}^{i_1}\Delta_F(q_{j})}
\label{4.72}
\end{equation}
because it has been taken into account  in the  term  $T^{n}_2$.}

\emph{ Moreover, by using the same arguments we presented  before for $H^{n+1}>0$ this term is positive (recurrently).} 
 
 \hspace{10cm}$\blacksquare$

\end{enumerate}
\end{Appendix}

\subsection{Reminders}
\begin{Appendix} \label{Ap.4.4}

\begin{enumerate}
\item{The equations of motion established in \cite{MM}}
\begin{definition}\label{def.4.1}
    \begin{equation}
 	H^2 (q,\Lambda)= - \frac{\Lambda}{\gamma+\rho}  
\lbrace\lbrack N^{(3)}_3H^4\rbrack
 -\Lambda\alpha
H^2 (q,\Lambda)\Delta_F(q) \rbrace +{(q^2+m^2)\gamma\over{\gamma+\rho}}
	\label{4.73}
 \end{equation}
 (Here $m>0$  and $\Lambda>0$ are the physical mass
 and coupling constant of the interaction model,
and $\alpha$, $\beta$, $\gamma$, are physically
 well defined quantities associated to this model,
the so called renormalization constants).
Moreover,
$$\forall\,  n \geq3, \, (q,\Lambda) \in 
{\cal E}^{4n}\times \Bbb R^{+*}$$
 \begin{equation}
 H^{n+1}(q,\Lambda) =  {1\over{\gamma+\rho}} \lbrace\,
 \lbrack A^{n+1}+ B^{n+1}
 + C^{n+1}\rbrack(q,\Lambda) + \Lambda 
\alpha	H^{n+1}(q,\Lambda)\Delta_F(q) \rbrace
 \label{4.74}
 \end{equation}
with:
\begin{equation}\begin{array}{l}
A^{n+1}(q,\Lambda) = - \Lambda
\lbrack N^{(n+2)}_3H^{n+3}\rbrack(q,\Lambda);\\
B^{n+1}(q,\Lambda) =
 - 3\Lambda\sum_{\varpi_n(J)} 
\lbrack N^{(j_2)}_2H^{j_{2}+2}
 N^{(j_1)}_1H^{j_{1}+1}\rbrack(q,\Lambda)
 \end{array}
 \label{4.75}
 \end{equation}
\begin{equation}C^{n+1}(q,\Lambda) = 
- 6\Lambda\sum_{I\in\varpi_n(3)}\prod_{l=1,2,3}
\lbrack N^{(i_l)}_1\vert H^{i_{l}+1}\vert\rbrack
(q_{i_{l}},\Lambda)
\label{tree.dec.}
 \end{equation}
Here the notations:
\begin{equation}
\lbrack N^{(n+2)}_3H^{n+3}\rbrack, \ 
  \lbrack N^{(j_2)}_2H^{j_{2}+2}\ 
 N^{(j_1)}_1H^{j_{1}+1}\rbrack \ \mbox{and} 
  \displaystyle\prod_{l=1,2,3}
\lbrack N^{(i_l)}_1H^{i_{l}+1}
\rbrack
\end{equation}
 represent the  $\Phi^4_4$ 
operations which have been 
introduced in the
 ``Renormalized
 G-Convolution  Product'' (R.G.C.P) 
 context of the references
 \cite{Br.MM}
\cite{MM5}.

 Briefly, the two
 loop  $\Phi^4_4$ - operation is defined by:
 \begin{equation}
\lbrack N^{(n+2)}_3H^{n+3}\rbrack	= \int R^{(3)}_G \lbrack\
 H^{n+3}\prod_{i=1,2,3}
\Delta_F(l_i)\ \rbrack d^{4}k_1d^{4}k_2 
\label{4.78}
\end{equation}
with $ R^{(3)}_G$ the corresponding 
renormalization operator
 for the two loop graph.
The analogous expression for the one loop  $\Phi^4_4$- operation is the following:
\begin{equation}
\lbrack N^{(j_2)}_2H^{j_{2}+2}
 N^{(j_1)}_1H^{j_{1}+1}\rbrack
= (H^{j_{1}+1}
\Delta_F)(q_{j_1})\,
\int R^{(2)}_G H^{j_{2}+2}\prod_{i=1}^{2}
\Delta_F(l_i)d^{4}k
\label{4.79}
\end{equation}

\end{definition}
The method is based on the proof of the existence 
and uniqueness of the solution of the 
corresponding infinite system of dynamical equations
 of motion verified by the sequence of the
Schwinger functions, i.e the connected, completely
 amputated with respect to the free propagator
Green's functions: 
\begin{equation}
 H =\{H^{n+1}\}_{n =2k+1, k \in\Bbb N}
 \end{equation}
 in the Euclidean $r$-dimensional
 momentum space, ${\cal E}^{rn}$
 (where $ 0\leq r\leq 4 $).  

\item{{The subset $\Phi_R \subset{\cal B}_R $}} \cite{MM7}
\begin{definition}\label{def.4.2}
We say that a sequence $H \in \mathcal{B}_R $ belongs to the 
subset $\Phi_R$, if the following properties are verified:
\begin{enumerate}
\item{} $\forall(q,\Lambda)\in \mathcal{E}^{4}\times]0,0.04]$    
\begin{equation}\begin{array}{l}
H^2 (q,\Lambda)= (q^2+m^2)(1+\delta_{1}(q,\Lambda)\Delta_F )\\
\mbox{with} \\
\delta_{1}(q,\Lambda)\Delta_F(q)|_{(q^2+m^2)= 0}=0\  \ \mbox{or}\   H^2 \Delta_F(q)|_{(q^2+m^2)= 0}=1\\
 \ \ \\
\mbox{and} \quad	 H^2_{min}(q)  \leq H^{2}(q,\Lambda) \leq H^{2}_ {(max)}(q,\Lambda) \\
\ \ \\
	 \mbox{with}\ \ \ H^{2}_ {(max)}(q,\Lambda)= \gamma_{max} (  ( q^{2}+m^2)+6\Lambda^2(q^2+m^2)^{\frac{\pi^2}{54}});\\
	   H^2_{min}(q)=q^2+m^2
	 \end{array}
	 \label{4.81}
  \end{equation}
\item{} 
For every  $n=2k+1, k\in\N^*$  the function $H^{n +1}$, belongs to the class 
$\mathcal{A}_{4n}^{(\alpha_{n}\beta_{n})}$
of Weinberg functions such that 
 $ \forall\  S\subset\mathcal{E}^{4n}$ the corresponding asymptotic 
indicatrices are given by:
\begin{equation}
	\alpha_{n}(S)=\left\{
	\begin{array}{l}
		-(n-3) \  \mbox{if } \ S\not\subset\mathcal{K}er\ \lambda_{n}  \\
		0 \ \  \mbox{if}\ S\subset\mathcal{K}er\ \lambda_{n}\\ 
	\beta_{n}(S)=n\beta_{1} \ \ \forall\  S\subset{\mathcal E}^{4n}\\
 \mbox{(with $\beta_{1}\in\N$  arbitrarily large)}
	\end{array}
	\right\}	
		\label{4.82}
\end{equation}
\item{} \label{def4.1;3} There is an increasing and bounded (with respect 
to $n$)  associated positive sequence:
$\{\delta_{n}(q,\Lambda)\}_{n=2k+1,k\in\N^*}$, 
  of splitting functions $ \in  \mathcal{D}$ which belong to the 
class   $\mathcal{A}_{(n)}^{(0, 0)}$ of Weinberg functions for every $n\geq 3 $ such
that $H$ is a  tree type sequence. More precisely:
\begin{description} 	
 	\item{i)}\ \
$\forall(q,\Lambda)\in \mathcal{E}^{12}\times]0,0.04] $
\begin{equation}\begin{array}{l}
	 H^4 (q\Lambda) = -\delta_{3}(q,\Lambda)
	 \prod_{\ell=1,2,3}H^{2}(q_{\ell},\Lambda)\Delta_{F}(q_{\ell})\\ 
	  \mbox{ with}\  \delta_{3}(q,\Lambda)\build\sim_{q\rightarrow\infty}^{}\Lambda\\ 
	\mbox{  For every finite fixed }  \ \tilde q\in\mathcal{E}^{12}\ \ \ 
	 \displaystyle{\lim_{\Lambda\to 0} 
 	\frac{\delta_{3}(\tilde q,\Lambda)}{\Lambda}}=6\\
	\ \ \\
  \mbox{and}\ \ \forall\Lambda\in]0,0.04]\quad
	\delta_{3,min}(\Lambda) \leq \delta_{3}(\tilde q, \Lambda)\leq
	\delta_{3,max}(\Lambda)  
	\end{array}
	\label{4.83}
\end{equation}
 	\item {ii)}
	For every $n=2k+1, k\geq 2$ and 
 	$\forall (q,\Lambda)\in\mathcal{E}^{12}\times]0,0.04] $:
\begin{equation}
	\begin{array}{l}
 	H^{n+1}(q,\Lambda)=\displaystyle{\frac{\delta_{n}(q,\Lambda)C^{n+1}(q,\Lambda)}
 	{3\Lambda n(n-1)}}\\
 \mbox{ with}\ \    \delta_{n}(q,\Lambda)\build\sim_{q\rightarrow\infty}^{}\Lambda\\
 \ \ \\
	\mbox{ For every finite fixed } \tilde q\in\mathcal{E}^{4n}\\
	\ \displaystyle{\lim_{\Lambda\to 0} 
 	\frac{\delta_{n}(\tilde q,\Lambda)}{\Lambda}\sim 3n(n-1)}\\
	\ \ \\
\mbox{and}\ \ \forall\Lambda\in]0,0.04],\quad
	\delta_{n,min}(\Lambda) \leq \delta_{n}(\tilde q, \Lambda)\leq
	\delta_{n,max} (\Lambda) 
	\end{array}
	 \label{splitting}
\end{equation}
 Here  $\{\delta_{n,min}\}$, (but not $\{\delta_{n,max}\}$)  are the splitting sequences lower  bounds of the  \emph{solution of the zero dimensional problem} ( cf. definition \ref{def.SplitBounds} of the reminders).
	\item {iii)}	
Moreover there is a finite number  $\delta_{\infty}\in\R^+$ a uniform bound 
independent of $H$ such that : 
\begin{equation}
	\displaystyle{\lim_{n\to \infty}\delta_{n}(\tilde q, \Lambda)}\leq\delta_{\infty}\ \  \ \forall\ \Lambda\in ]0,\ 0.04]
	\label{4.85}
\end{equation}
 \end{description}
\item {} 
The \emph{renormalization functions} $a,\rho$ and $\gamma$, appearing in the 
    definition of $\mathcal{M}$ are well defined real analytic functions of 
    $q^2$ and $\Lambda$, and yield at the limits $(q^{2}+m^{2})=0$\  and \ $q=0$  the physical conditions of renormalization  required by the two-point and four
    point functions: 
    \begin{equation}\begin{array}{l}  
    	a (q, \Lambda)= [N_{3}^{(3)}H^4(q, \Lambda)]  \ \ \mbox{and} \ \  \tilde a(\Lambda)=[N_{3}^{(3)}H^4(q, \Lambda)]|_{(q^2+m^2)= 0}\\ 
\mbox{with:}	\ a_{min}(\Lambda)\leq \tilde a(\Lambda) \leq  a_{max}(\Lambda)
	\end{array}
	\label{4.86}
\end{equation}
\begin{equation}\begin{array}{l}
   	\rho(q, \Lambda)= \left[
  	\displaystyle{\frac{\partial }{\partial q^{2}}}[N_{3}^{(3)}H^4(q\Lambda)]	\right], \ \ \mbox{and} \ \  \tilde \rho (\Lambda)= \rho(q, \Lambda)|_{(q^2+m^2)= 0} \\
	\mbox{with:}\ \ \  \rho_{min}(\Lambda)\leq \tilde \rho (\Lambda)\leq\rho_{max}(\Lambda)
	\end{array}
	 \label{4.87}
\end{equation} 
\begin{equation}\begin{array}{l}
 \gamma(q,\Lambda)=\left[\displaystyle{\frac{-6\Lambda \prod_{l=1,2,3}H^{2}(q_l)\Delta_{F}(q_l)}{H^{4}(q)}}\right]\\  \ \\
\mbox{and}\ \tilde\gamma(\Lambda)  =
		\ \gamma(q,\Lambda)|_{q=0} \quad
	\mbox{with} \ \ \gamma_{min}(\Lambda)\leq \tilde\gamma(\Lambda)\leq\gamma_{max}(\Lambda)\\ 
	\end{array}  
	\label{4.88}
\end{equation}
\end{enumerate}
\end{definition}
\item{}
\begin{definition}\label{def.SplitBounds} {The upper and lower bounds of the splitting sequences    and of the renormalization parameters} (cf.\cite{MM7} )
\begin{equation}\begin{array}{l}
\forall \Lambda \in\ ]0,0.04]\\
\    \delta_{3, max}(\Lambda) =\displaystyle{\frac{6\Lambda}{ 1+\rho_0+\Lambda|a_0| +6d_0}};\qquad
 \delta_{3, min}(\Lambda) = \displaystyle{{6\Lambda\over 1+9\Lambda(1+6\Lambda^2)}}\\
\mbox{and}   \ \  \forall   n \geq 5 \\
 \delta_{n, max}(\Lambda) = \displaystyle{\frac{3\Lambda  n(n-1)}{ 1+\rho_0+\Lambda|a_0| +n(n-1)d_0}}\\
 \mbox{with:}\\
   \quad  a_0 = -\delta_{3,min}[N_{3}\tilde ]_{q^2+m^2=0};\quad 
	 \rho_{0}= \Lambda\delta_{3,min}\lbrack \displaystyle{\frac{\partial}{\partial q^{2}}[N_{3}\tilde ]\rbrack_{q^2+m^2=0}}\\
 \mbox{and}\\ 
\delta_{n, min}(\Lambda ) =\displaystyle{{3\Lambda\ n(n-1)\over \gamma_{max}+\rho_{max}+\Lambda|a_{max}| +3\Lambda  n\ (n-1)} 
 }\\
\mbox{with}\\
\gamma_{max}=1+9\Lambda (1+6\Lambda^2),\gamma_{min}=1,\  \rho_{max}=6\Lambda^2 \displaystyle{\frac{\partial}{\partial q^{2}}}[N_{3}\tilde ]_{q^2+m^2=0}\ \ \ \mbox{and} \\
|a_{max}|=  6\Lambda[N_{3}\tilde ]_{q^2+m^2=0}\\
\end{array}
\label{4.89}
\end{equation}
\end{definition}
 
 \item{The signs and bounds }
 
 The following properties have been established   in \cite{MM7}\cite{MM8} at every order of the $\Phi_4^4 $-iteration consequently, the sequence $\{H\}$ solution of the contractive mapping ${\cal M}^*$ also  verifies the following: 
\begin{proposition}  \label{prop.5.1}\ 
 $  \forall \Lambda \in\ ]0,0.04]$
\begin{itemize}
\item[i)] 
$ \forall   q\in \mathcal{E}_{(q)}^{4}$
\begin{equation}\begin{array}{l}
 H^2(q,\Lambda)>0, \qquad 
   H^2_{min}(q,\Lambda)\leq  H^2(q,\Lambda) \leq  H^2_{(max)}(q,\Lambda)\\
   \ \ \\
    \mbox{with}\ \  H^2_{min}(q,\Lambda)=q^2+m^2\\
 \mbox{and}\\
   H^2_{(max)}(q,\Lambda)=\gamma_{max}[(q^2+m^2)+6\Lambda^2(q^2+m^2)^{\pi^2/54} ]\\
\end{array}
\label{4.90}
\end{equation}
\item[ii)] The 
global term  $(`\Phi_4^{4}$ operation'')
\begin{equation} C^{n+1}(q,\Lambda) = - 6\Lambda\displaystyle{\sum_{\varpi_n(I)}\prod_{l=1,2,3}}
\lbrack N^{(i_l)}_1H^{i_{l}+1}\rbrack (q_{i_{l}},\Lambda)
\label{4.91}
\end{equation} 
    verifies the following properties:
\begin{itemize}
\item[a.] The ``good sign'' property:
\begin{equation}\forall \ n=2k+1\  (k\geq 1)\ \  C ^{n+1}=(-1)^{\frac{n-1}{2}}|C^{n+1}| 
\label{goodsign}
\end{equation}
\item[b.] It  is a R.$\Phi$.C. (cf.\cite{MM7} consequently it verifies Euclidean invariance and linear axiomatic quantum field theory properties.
\item[c.]
For every $n=2k+1, k\geq 1$  the function $ C^{n+1}(q, \Lambda)$ , belongs to the class 
$\mathcal{A}_{4n}^{(\alpha_{n}\beta_{(n )})}$
of Weinberg functions such that 
 $ \forall\  S\subset\mathcal{E}_{(q)}^{4n}$ the corresponding asymptotic 
indicatrices are given by:
\begin{equation}
	\alpha_{n}(S)=\left\{
	\begin{array}{rl}
		-(n -3),  & \mbox{if } \ S\not\subset\mathcal{K}er\ \lambda_{n}  \\
		0 & \mbox{if }\ S\subset\mathcal{K}er\ \lambda_{n} 
	\end{array}
	\right\}
	\label{4.93}
\end{equation}
\begin{equation}
	\beta_{(n, )}= \beta_{(1 )}n\ \ \ \forall\  S\subset\mathcal{E}_{(q)}^{4n}
		\label{4.94}
\end{equation}
\item[d)] For every $n=2k+1, k\geq1$  
\begin{equation} \begin{array} {l}      \vert  C^{n+1}_{min}(q, \Lambda)\vert\leq	\vert  C ^{n+1}(q,\Lambda)\vert\leq \vert  C^{n+1}_{ max}(q, \Lambda)|\\
\mbox{with}: \\
|C^{n+1}_{max}(q,\Lambda)|= 3\Lambda  n(n-1) \mathcal{T}_{n} |H^{n-1}(q_{{\scriptscriptstyle (n-2)}} )|\displaystyle{\prod_{l=2,3} (H^2\Delta_F)(q_{{l}} )}\\
 \vert  C^{n+1}_{min}(q, \Lambda)\vert=3\Lambda  n(n-1) \tilde {\mathcal T} _{n} |H^{n-1}(q_{{\scriptscriptstyle (n-2)}}) |\displaystyle{\prod_{l=2,3}(H^2\Delta_F)(q_{{l}} )}
 \end{array}
\label{4.95}
\end{equation}

\begin{remark}\

Notice that in the last formula we take  into account  the result
of ref. \cite[c]{MM1} on the number $\mathcal{T}_{n}$ (and $\tilde {\mathcal T} _{n}$)
 of different partitions inside the tree terms. Precisely:
\begin{equation}\begin{array}{l}
\mbox{for}\  n=3, n=5 \ \mathcal{T}_{n}=1 \\
\mbox{and}, \\
 \forall n\ \geq 7\ \  \mathcal{T}_{n}=[\frac{(n-3)^{2}}{48}]+[\frac{(n-3)}{3}]+1\\
    \mbox{(where $[.]$ means integer part )}
    \ \ \\
    \mbox{and} \quad \tilde {\mathcal T} _{n}=[\frac{(n-3)^{2}}{48}]+[\frac{(n-3)}{3}]
    \end{array}
    \label{4.96}
\end{equation}
\label{rem.4.2}
\end{remark}
\end{itemize}
\item[iii)]
\begin{equation}\begin{array}{l}
 \forall \ n=2k+1\  (k\geq 1)\ \  H ^{n+1}=(-1)^{\frac{n-1}{2}}|H^{n+1}| 
\end{array}
\label{4.97}
\end{equation}
\item[iv)]
\begin{equation}
 \forall \ n=2k+1\  (k\geq 1) \quad 	\vert  H^{n+1}_{min} \vert\leq	\vert	H ^{n+1} \vert\leq \vert H^{n+1}_{max}\vert
\label{4.98}
\end{equation}
\end{itemize}
Here $H^{n+1}_{max}$ is  defined as follows: 
\begin{equation}
\vert H^{4}_{max}\vert=\delta_{3,max}\displaystyle{\prod_{l=1,2,3}  H^2(q_{{l}},\Lambda)\Delta_F(q_{{l}} )}
\label{4.99}
\end{equation}
Then recurrently $\forall n=2k+1-k\geq 2$  and by  using the preceding definitions of  $\vert  C^{n+1}_{max}(q, \Lambda)\vert$ and $\vert  C^{n+1}_{min}(q, \Lambda)\vert:$ we obtain the bounds:
\begin{equation}
\begin{array}{l}
 |H^{n+1}_{(max)}(q_{{\scriptscriptstyle (n)}})|\equiv \delta_{n,max} \mathcal{T}_{n}\Delta_F(\sum_{i=1}^{n-2}q_i) |H^{n-1}_{max}(q_{{\scriptscriptstyle (n-2)}})|\displaystyle{\prod_{l=2,3}  H^2(q_{{l}})\Delta_F(q_{{l}})}\\
\\
|H^{4}_{min}(q_{{\scriptscriptstyle (3)}})|= \delta_{3,min} \displaystyle{\prod_{l=1,2,3}  H^2(q_{{l}},\Lambda)\Delta_F(q_{{l}} )}\\
|H^{n+1}_{min}(q_{{\scriptscriptstyle (n)}})|= \delta_{n,min} \tilde {\mathcal T} _{n}\Delta_F(\sum_{i=1}^{n-2}q_i)|H^{n-1}_{min}(q_{{\scriptscriptstyle (n-2)}})|\displaystyle{\prod_{l=2,3}  H^2(q_{{l}},\Lambda)\Delta_F(q_{{l}} )}\\
\end{array}
\label{4.100}
\end{equation}
\label{prop.5.1}
\end{proposition}
\end{enumerate}
\label{App.5.3} 
\end{Appendix}
\end{document}